\documentclass[a4paper,11pt]{article}

\usepackage{cite}
\usepackage[T1]{fontenc}

\usepackage{amssymb,amsmath}

\usepackage{framed}
\usepackage{mdframed}

\usepackage{float}
\allowdisplaybreaks

\begin{document}
\vspace{5mm}

\def\be{\begin{eqnarray}}
\def\ee{\end{eqnarray}}

\begin{titlepage}
\thispagestyle{empty}
\begin{flushright}
{\ }	
\end{flushright}
				
\vspace{35pt}
				
\begin{center}
	{\Large{\bf Liberated ${\cal N}=1$  Supergravity }}
									
	\vspace{30pt}
							
	{Fotis Farakos${}^{1}$, Alex Kehagias${}^{2}$ and 
	Antonio Riotto${}^{3}$}
							
	\vspace{15pt}
							
	{
{\ }$^{1}${\normalsize {\it KU Leuven, Institute for Theoretical Physics, \\
			Celestijnenlaan 200D, B-3001 Leuven, Belgium}}
		\vskip.07in
{\ }
{\ }$^{2}${\normalsize {\it Physics Division, National Technical University of Athens, 
			\\ \it 15780 Zografou Campus, Athens, Greece }}
			\vskip.07in
{\ }
			$^{3}${\normalsize {\it D\'epartement de Physique Th\'eorique and Centre for Astroparticle Physics (CAP), \\
			Universit\'e de Gen\`eve, 24 quai E. Ansermet, CH-1211 Geneva, Switzerland	}}		
			}

\vspace{40pt}
								
{ABSTRACT} 
\end{center}

We discuss a new interaction for chiral models in four-dimensional ${\cal N}=1$  supergavity. 
It contains a new arbitrary function in addition to the K\"ahler potential and superpotential. 
Its  features include  linearly realized off-shell supersymmetry, K\"ahler-Weyl invariance and  broken supersymmetry. 
The corresponding scalar potential is augmented  by the arbitrary function  which allows freedom in constructing low-energy phenomenological models and inflationary models rooted in supergravity.

\vspace{10pt}
			
\bigskip
			
\end{titlepage}

\numberwithin{equation}{section}

\baselineskip 6 mm

\tableofcontents

\bigskip

\def\be{\begin{eqnarray}}
\def\ee{\end{eqnarray}}

\def\ls{\left[}
\def\rs{\right]}
\def\lc{\left\{}
\def\rc{\right\}}

\def\p{\partial}

\def\S{\Sigma}

\def\s{\sigma}

\def\O{\Omega}

\def\a{\alpha}
\def\b{\beta}
\def\g{\gamma}

\def\ad{{\dot \alpha}}
\def\bd{{\dot \beta}}
\def\gd{{\dot \gamma}}

\def\nn{\nonumber}

\allowbreak

\def\thefootnote{\arabic{footnote}}
\setcounter{footnote}{0}

\addtocounter{page}{1}

\baselineskip 6 mm

\section{Introduction}

Four-dimensional ${\cal N}=1$ supergravity is highly motivated for phenomenological purposes as it provides an  
appropriate setup to describe low-energy effective field theories originating from string theory. 
The main focus in this direction has been the study of supergravity theories where 
supersymmetry can in principle be restored within the validity of the supergravity theory. 
This procedure requires the study of supergravity theories 
with broken or unbroken supersymmetry 
where there always exists a smooth limit to the restoration of supersymmetry within the regime of validity of the effective supergravity theory.

However, 
this is not an essential criterion that any low-energy supergravity theory originating from string theory has to satisfy, 
as there exist known constructions containing (anti)branes, 
where supersymmetry might not be restored within the supergravity limit. 
Notable examples are the brane supersymmetry breaking setup \cite{Antoniadis:1999xk,Angelantonj:1999jh,Aldazabal:1999jr,Dudas:2000nv,Angelantonj:1999ms,Pradisi:2001yv,Mourad:2017rrl}, 
and the KKLT scenario \cite{Kachru:2003aw,Kachru:2003sx,Bergshoeff:2015jxa}. 
Indeed a paradigm shift was considered only recently, 
where models of supergravity with non-linear realizations have been investigated, 
where supersymmetry is not allowed to be restored \cite{Dudas:2015eha,Bergshoeff:2015tra,Hasegawa:2015bza,Bandos:2015xnf,Cribiori:2016qif,Bandos:2016xyu,DallAgata:2015pdd}. 
Constrained superfields in supergravity were however known earlier, for example since the work of  \cite{Lindstrom:1979kq,Samuel:1982uh}, 
and the contrast of such theories to standard supergravity is striking. 
In particular new forms of the scalar potential are allowed which can easily describe inflation or a KKLT-type uplift \cite{Antoniadis:2014oya,Ferrara:2014kva,Kallosh:2014via,DallAgata:2014qsj,McDonough:2016der,Kallosh:2017wnt}.

Following this line of thought, 
new terms have been constructed in supergravity where effects similar to the non-linear realizations have been achieved, 
but the theory has off-shell supersymmetry linearly realized. 
In particular in \cite{Cribiori:2017laj} 
an uplift usually attributed to non-linear realizations (anti-D3 branes) has been constructed  where supersymmetry is linearly realized, albeit spontaneously broken by the auxiliary field of the vector multiplet. 
In \cite{Antoniadis:2018oeh} K\"ahler invariance is restored and a constant uplift is described. 
These novel results indicate that all the effects that have been studied with constrained superfields and non-linear realizations of supersymmetry 
can be instead studied with supersymmetry linearly realized. 
Let us note in passing that models for inflation utilizing the setup of \cite{Cribiori:2017laj} 
have been studied in \cite{Aldabergenov:2017hvp,Antoniadis:2018cpq}.\footnote{Theories with similar properties have been also investigated in \cite{Kuzenko:2017zla,Kuzenko:2018jlz}.}

In this work we do one further step towards this new direction. 
We study chiral models coupled to supergravity and we show that whenever supersymmetry is spontaneously broken 
there exists a deformation of the scalar potential of the form 
\be
\label{pot-intro}
{\cal V} = {\cal V}_\text{SUGRA} + {\cal U}(A^I, \overline A^{\overline J})  \, . 
\ee
This deformation is induced by off-shell linear realizations of supersymmetry, 
and by construction also respects the K\"ahler invariance of the standard supergravity theory. 
The consistency of this new contribution requires that supersymmetry is broken by at least one of the auxiliary fields of the 
standard chiral multiplets, 
and the positivity of the K\"ahler metric of the scalar manifold. 
For gauged chiral models the consistency of the new term can be also guaranteed if supersymmetry is broken by the gauge sector. 
In other words the consistency requirements for the new term to be well-defined are absolutely minimal, 
they are model independent, and they are satisfied under 
any circumstance where four-dimensional ${\cal N}=1$ supersymmetry is broken spontaneously.

The addition of the new interaction term relaxes the form of the scalar potential giving more freedom to its structure. This has  consequences for many applications, for instance in low-energy  phenomenological models build up out of supergravity and in models of primordial inflation avoiding the so-called $\eta$-problem \cite{LR}. For this reason, we dub this
extension ``Liberated Supergravity''. 

The paper is organized as follows. In section 2 we review the chiral models of supergravity, while in section 3 we present the concept of 
liberated ${\cal N}=1$ supergravity, presenting its equivalent formulation with constrained superfields in section 4. Section 5 contains the particular case of a single superfield and  section 6 our conclusions and outlook.

\section{Chiral models in supergravity}

Let us quickly review some of the basic properties of the chiral multiplets coupled to minimal ${\cal N}=1$ supergravity. 
Our intention is to point out   some aspects of the theory which are of relevance in the following sections. 
The subject is now standard \cite{West:1990tg,Wess:1992cp,Buchbinder:1998qv,FvP} and we will follow here \cite{Wess:1992cp}.

In the old-minimal formulation of supergravity the component fields of the supergravity multiplet are the vielbein $e_m^{\ a}$ which describes gravity, 
and its superpartner the gravitino $\psi_m^{\ \alpha}$ which is a spin-3/2 fermion. 
The auxiliary fields of the supergravity multiplet are the complex scalar $M$ 
and the real vector $b_a$ \cite{Stelle:1978ye,Ferrara:1978em,Wess:1978ns}. 
The local supersymmetry transformations of the supergravity multiplet are 
\be
\label{ddsugra} 
\begin{aligned}
\delta e_m^{\ a} = & \, i \left( \psi_m \sigma^a \overline \xi - \xi \sigma^a \overline \psi_m \right) \, , 
\\
\delta \psi_m^{\ \alpha} = & - 2 D_m \xi^\alpha + i e_m^{\ c} \lc \frac13 M (\epsilon \sigma_c\overline \xi)^\alpha 
+ b_c \xi^\alpha 
+\frac13 b^d (\xi \sigma_d \overline \sigma_c)^\alpha 
\rc \, , 
\\
\delta M= & - \xi ( \sigma^a \overline \sigma^b \psi_{ab} 
+ i b^a \psi_a 
- i \sigma^a \overline \psi_a M ) \, , 
\\ 
\delta b_{\alpha \dot \alpha} = & \, \xi^\delta \left\{ \frac{3}{4} \overline \psi^{\ \dot \gamma}_{\alpha \  \delta \dot \gamma \dot \alpha} 
+ \frac{1}{4} \epsilon_{\delta \alpha}  \overline \psi^{\gamma \dot \gamma}_{\ \ \gamma \dot \alpha \dot \gamma} 
-\frac{i}{2} \overline M \psi_{\alpha \dot \alpha \delta}  
+\frac{i}{4} ( \overline \psi_{\alpha \dot \rho}^{\ \ \dot \rho} b_{\delta \dot \alpha} 
+\overline \psi_{\delta \dot \rho}^{\ \ \dot \rho} b_{\alpha \dot \alpha} 
-\overline \psi _{\delta \ \dot \alpha}^{\dot \rho} b_{\alpha \dot \rho} )  \right\} + {\rm c.c}.
\end{aligned}
\ee 
For the gravitino we have $\psi_{nm}^{\ \ \ \alpha} = D_n \psi_m^{\ \alpha} - D_m \psi_n^{\ \alpha}$ 
where $D_m \psi_{n \alpha} = \partial_m \psi_{n \alpha} - \omega_{m \alpha}^{\ \ \ \beta} \psi_{n \beta}$, 
and for the supersymmetry parameter we have $D_m \xi_\alpha = \partial_m \xi_\alpha - \omega_{m \alpha}^{\ \ \ \beta} \xi_\beta$.

Let us consider a set of chiral superfields 
\be
\Phi^I = A^I + \sqrt 2 \Theta^\alpha \chi_\alpha^I + \Theta^2 F^I \, . 
\ee  
The local supersymmetry transformations are 
\be
\begin{aligned}
\label{ddphi} 
\delta A^I &= - \sqrt{2} \xi \chi^I \, , 
\\
\delta \chi_\alpha^I &= - \sqrt{2} F^I \xi_\alpha 
- i \sqrt{2} \s_{\alpha \dot \alpha}^{a}  \overline \xi^{\dot \alpha} \hat D_a A^I \, , 
\\
\delta F^I &=  - \frac{\sqrt{2}}{3} \overline M \xi \chi^I 
- \bar\xi^{\dot \alpha} ( i \sqrt 2 \hat D_{\alpha \dot \alpha} \chi^{I \alpha} 
- \frac{\sqrt 2}{6} b_{\alpha \dot \alpha} \chi^{I \alpha} ) \, , 
\end{aligned}
\ee
where we have made use of the supercovariant derivatives
\be
\hat D_a A 
=   e^{\ m}_{a} 
\left( \p_m A -\frac{1}{\sqrt{2}} \psi_{m}^{ \a} \chi_\a \right) \, , \quad 
\hat D_a \chi_\alpha 
= e^{\ m}_{a} 
\left( 
D_m \chi_\alpha 
-\frac{1}{\sqrt{2}} \psi_{m \a} F 
- \frac{i}{\sqrt{2}} \overline \psi_{m }^{\ \dot \b} \hat D_{\a \dot \b} A
\right) \, . 
\ee 
We couple the $\Phi^I$ to standard supergravity via\footnote{We use $\int {\rm d}^4 \theta \, E \, \Omega = \int {\rm d}^2 \Theta \, 2 {\cal E} \, 
\ls -\frac18 (\overline {\cal D}^2 - 8 {\cal R}) \, \O \rs + {\rm c.c}.$ up to boundary terms. In \cite{Wess:1992cp} 
the explicit expressions for $2 {\cal E}$ and ${\cal R}$ can be found.} 
\be
\label{gen1}
{\cal L}_0 =
\int {\rm d}^4 \theta \, E \, \Omega(\Phi^I, \overline \Phi^{\overline J} ) + 
\left(  \int {\rm d}^2 \Theta \, 2 {\cal E} \, W(\Phi^I) + {\rm c.c}.  \right) \, . 
\ee
Here $\Omega$ is a real function which is related to the K\"ahler potential as 
\be
K = -3 \, \text{log}( - \O /3) \, , 
\ee
and $W$ is a holomorphic function of the chiral superfields. 
Standard ${\cal N}=1$ supergravity is invariant under super-Weyl-K\"ahler 
transformations with chiral superfield parameter $\Sigma(\Phi^I)$, under which the   K\"ahler potential $K$ and the superpotential transform as 
 \be
\label{KSS}
K \rightarrow K + 6 \Sigma + 6 \overline \Sigma \, , \quad  W \rightarrow W \text{e}^{- 6 \Sigma}.
\ee
The above transformations are symmetries of the superspace Lagrangian (\ref{gen1}) if they are accompanied by compensating transformations of 
the superspace measures 
\be
\label{SWI}
{\rm d}^4 \theta \, E \rightarrow {\rm d}^4 \theta \, E \, \text{e}^{2 \Sigma + 2 \overline \Sigma} 
\, , \quad  
{\rm d}^2 \Theta \,  2{\cal E} \rightarrow {\rm d}^2 \Theta \,  2{\cal E} \, \text{e}^{6 \Sigma} \, , 
\ee
where also the chiral projection transforms under the super-Weyl-K\"ahler as (see for example \cite{Buchbinder:1998qv}) 
\be
\label{WPR} 
- \frac14 \left( \overline{\cal D}^2 - 8 {\cal R} \right) \rightarrow 
- \frac14 \left( \overline{\cal D}^2 - 8 {\cal R} \right) \text{e}^{-4 \Sigma + 2 \overline \Sigma} \, . 
\ee 
Once the  auxiliary fields are eliminated and a Weyl rescaling is performed in  order to write the theory in the Einstein frame, 
the bosonic sector of the standard ${\cal N}=1$  supergravity \eqref{gen1}  takes the form 
\be
\label{BOS}
e^{-1}  {\cal L}_0 |_\text{bosonic} = - \frac12 R - g_{I \overline J} \p_m A^I \p^m \overline A^{\overline J}  - {\cal V} \, , 
\ee 
where 
\be
{\cal V} = \text{e}^K \left[ D_{\overline J} \overline W g^{\overline J I} D_I W  - 3 W \overline W \right] \, , 
\ee
and as usual we have $D_I W = W_I + K_I W$,  $W_I=\partial W/\partial A^I$ and $K_I=\partial K/\partial A^I$. 
The K\"ahler metric is defined as $g_{I \overline J} = K_{I \overline J}$ 
and $g^{\overline J I}$ is its inverse. 
In this standard setup the complex  gravitino mass is given by 
\be
m_{3/2} = \text{e}^{K/2} W \, . 
\ee
Notice that the positive contribution to the scalar potential, 
which is essentially related to the breaking of supersymmetry, is sourced by the term
\be
D_{\overline J} \overline W g^{\overline J I} D_I W \sim F^I g_{I \overline J} \overline F^{\overline J} \, , 
\ee
whereas the negative contribution is related to the gravitino mass, 
and has the form $-3 |m_{3/2}|^2$. 
We remind  that on-shell, before Weyl rescaling, we have for the bosonic contributions to the matter auxiliary fields 
\begin{eqnarray}
F^I = -e^{ K/3} g^{I \overline J} D_{\overline J} \overline W. 
\end{eqnarray}
It is therefore important to realize that in any situation where supersymmetry is spontaneously broken 
we will have the model independent property 
\be
\label{SUSYB}
\text{Broken supersymmetry}: \quad \langle F^I g_{I \overline J} \overline F^{\overline J} \rangle \ne 0 \, . 
\ee
As we will see,  
\eqref{SUSYB} is the only consistency condition for the new Lagrangian 
 we will  propose next to be well-defined. 
In addition, we should also mention that \eqref{SUSYB} is 
satisfied whenever at least one of the vevs of the auxiliary fields $\langle F^{I} \rangle$ 
is non-vanishing due to the positivity of the K\"ahler metric.  
Let us finally present the bosonic contribution to the on-shell value of the supergravity scalar auxiliary field $M$, 
which before Weyl rescaling is given by 
\be
\label{Mcomp}
M = - 3 W \text{e}^{K/3} - K_{\overline J} \overline F^{\overline J} \, . 
\ee

\section{Liberated ${\cal N}=1$ supergravity}

When supersymmetry is non-linearly realized 
one can introduce a variety of new terms which deform the scalar potential 
in ways not allowed by the standard supergravity with linearly realized supersymmetry. 
In \cite{Cribiori:2017laj} a new coupling has been introduced where 
supersymmetry is linearly realized off-shell, but does however generate the uplift which is usually attributed to non-linear realizations. 
Therefore in such setup the uplift can in fact be described by linear supersymmetry, 
which is nevertheless broken by the vev of the auxiliary field of an abelian vector multiplet. 
The constructions in \cite{Cribiori:2017laj,Antoniadis:2018oeh} pave the way for a novel understanding of spontaneous supersymmetry breaking by linearly realized supersymmetry, and our work is in the spirit of this approach.

In this section we will work with the chiral multiplets $\Phi^I$ coupled to supergravity. 
We present the superspace formula and the component form (up to two fermions) of the Lagrangian term responsible for inducing the 
generic contribution ${\cal U}(A^I , \overline A^{\overline J})$ in the scalar potential \eqref{pot-intro}. 
We study the coupling of the new term to standard supergravity and we discuss its properties. 
Our construction here is similar to the one in \cite{Cribiori:2017laj}, 
however instead of introducing a Fayet--Iliopoulos term we will use the method presented there 
to introduce a direct uplift in the scalar potential, 
thus liberating it from its standard form. 
In the bulk of this section we do not consider the coupling to any gauge multiplet, 
however we comment on this extension in the end of the section.

\subsection{The new term} 

In order to construct our new supergravity Lagrangian, we will assume that  K\"ahler invariance is still a good symmetry 
and it is respected \cite{Cecotti:1986qw}. 
Since the K\"ahler transformation of the K\"ahler potential $K(\Phi^I, \overline \Phi^{\overline J} )$ 
in \eqref{KSS} is like  an effective abelian gauge transformation of a vector multiplet, 
it is clear that in order to maintain K\"ahler invariance, 
we should employ the field strength  of  $K(\Phi^I, \overline \Phi^{\overline J} )$. 
Therefore, we  define the spinor chiral superfield  
\be
{\cal W}_\alpha(K) = - \frac14 \left( \overline{\cal D}^2 - 8 {\cal R} \right) {\cal D}_\alpha K \, , 
\ee 
which is clearly invariant under the K\"ahler transformation \eqref{KSS}.
It has lowest component field the fermion 
\be
\label{eta}
\eta_\alpha \equiv i {\cal W}_\alpha | = i \sqrt 2 K_{I \overline J} \overline F^{\overline J} \chi_\alpha^I 
- \frac{i}{\sqrt 2} K_{I \overline L \, \overline J}\, \overline \chi^{\overline L} \overline \chi^{\overline J} \chi_\alpha^I 
+ \sqrt 2 K_{I \overline J} \, \sigma_{\alpha \dot \rho}^a \, \overline \chi^{\dot \rho \overline J} e_a^{\ m} \hat D_m A^I \, . 
\ee 
Notice that the structure of this fermion is very similar to the goldstino appearing in standard supergravity:  
$K_{I \overline J} \overline F^{\overline J} \chi_\alpha^I$. 
Of course this is not by chance since here the new terms we will introduce 
are allowed if and only if supersymmetry is spontaneously broken by the chiral superfields. 
Since ${\cal W}_\alpha(K)$ is a spinor chiral superfield (the various components of which can be found in formula (14.15) of \cite{FvP}), 
the only pure bosonic part  is in the component
\be
\label{dw}
 {\cal D}_\beta {\cal W}_\alpha(K) |_\text{bosons} = 
-2 \epsilon_{\beta \alpha} g_{I \overline J} F^I \overline F^{\overline J} 
+ 2 \left( \sigma^b_{\beta \dot \beta} \epsilon^{\dot \beta \dot \alpha} \sigma^a_{\alpha \dot \alpha} \right) e_{a}^{\ m} e_{b}^{\ n} 
g_{I \overline J} \,  \partial_m A^I \partial_n \overline A^{\overline J}
 \equiv  {\cal F}_{\beta \alpha}\, , 
\ee
where we define the composite bosonic field ${\cal F}_{\beta \alpha}$ for later use. 
Notice that when the condition \eqref{SUSYB} holds, 
we will have 
\be
\label{COND}
\langle {\cal D}^2 {\cal W}^2 (K) \big| \rangle 
= -2 \langle {\cal F}^{\alpha \beta} {\cal F}_{\alpha \beta}  \rangle 
=  - 16 \left( \langle g_{I \overline J} F^I \overline F^{\overline J} \rangle \right)^2  \ne 0 \, , 
\ee
which means that the inverse  of ${\cal D}^2 {\cal W}^2 (K)$ exists and it is $1/{\cal D}^2 {\cal W}^2 (K)$. 
Now we are ready to introduce the term 
\be
\boxed{
\label{LNEW}
{\cal L}_\text{NEW} = - 16 \, \int {\rm d}^4 \theta \, E \, \text{e}^{-2K/3} \, \frac{
{\cal W}^2(K) \overline {\cal W}^2(K) 
}{{\cal D}^2 {\cal W}^2(K) \overline {\cal D}^2 \overline {\cal W}^2(K)} \ {\cal U}(\Phi^I , \overline \Phi^{\overline J})  \, , }
\ee
where ${\cal U}(\Phi^I , \overline \Phi^{\overline J})$ is a general real function of the chiral superfields $\Phi^I$ and $\overline \Phi^{\overline J}$. 
To maintain K\"ahler-Weyl invariance as in the standard supergravity then ${\cal U}$ should be invariant under such transformations. 
Indeed, using the fact that under super-Weyl transformations 
we have ${\cal D}_\alpha K \rightarrow \exp\{ \Sigma - 2 \overline \Sigma\} {\cal D}_\alpha K$, 
and taking into account \eqref{WPR}, 
we see that  the ${\cal W}_\alpha(K)$ superfield changes under K\"ahler-Weyl transformations as 
\be
\label{WKW}
{\cal W}_\alpha(K) \rightarrow {\cal W}_\alpha(K) \, \text{e}^{-3 \Sigma} \, . 
\ee
This leads to 
\be
\frac{
{\cal W}^2(K) \overline {\cal W}^2(K) 
}{{\cal D}^2 {\cal W}^2(K) \overline {\cal D}^2 \overline {\cal W}^2(K)} 
\rightarrow 
\frac{
{\cal W}^2(K) \overline {\cal W}^2(K) 
}{{\cal D}^2 {\cal W}^2(K) \overline {\cal D}^2 \overline {\cal W}^2(K)} 
\, \text{e}^{2 \Sigma + 2 \overline \Sigma} \, , 
\ee
therefore \eqref{LNEW} is K\"ahler-Weyl invariant.

To built some intuition about the properties of this term we notice that we can recast the new Lagrangian 
in terms of the spinor goldstino superfield studied in \cite{Samuel:1982uh}. 
To this end we can define the composite spinor superfield 
\be
\label{WL}
\Gamma_\alpha \equiv -2 \, \frac{{\cal D}_\alpha {\cal W}^2(K)}{{\cal D}^2 {\cal W}^2(K)} \, , 
\ee 
which satisfies 
\be
\label{SWR}
\begin{aligned}
{\cal D}_\alpha \Gamma_\beta  &= \epsilon_{\beta \alpha} \left( 1 - 2 \, \Gamma^2 {\cal R} \right)  \, , 
\\
\overline{\cal D}^{\dot \beta} \Gamma^\alpha &= 2 i \, \left( \overline \s^a \, \Gamma \right)^{\dot \beta} \, {\cal D}_a \Gamma^\alpha 
+ \frac12 \, \Gamma^2 {\cal G}^{\dot \beta \alpha} \, .  
\end{aligned}
\ee
The superfield ${\cal G}^{\dot \beta \alpha}$ is defined for example in \cite{Wess:1992cp}. 
From \eqref{WL} we have that 
\be
\Gamma^2 \overline \Gamma^2 \equiv 16 \frac{
{\cal W}^2(K) \overline {\cal W}^2(K) 
}{{\cal D}^2 {\cal W}^2(K) \overline {\cal D}^2 \overline {\cal W}^2(K)} \, ,
\ee
which can be used to write the new Lagrangian term \eqref{LNEW} in a much more compact notation, 
namely 
\be
\label{LNSW}
{\cal L}_\text{NEW} = - \int {\rm d}^4 \theta \, E \, \text{e}^{-2K/3} \, \Gamma^2 \overline \Gamma^2 \ {\cal U}  \, . 
\ee
Here $\Gamma$ is not an independent constrained superfield, rather it is the composite superfield given by \eqref{WL}. 
Now one can directly see from the first formula in \eqref{SWR} that \eqref{LNSW} once expanded in component fields will  start 
with a bosonic term of the form $\text{e}^{-2K/3}\,{\cal U}$, 
and the rest of the terms will be fermionic.  
In particular, if we  denote 
\be 
\label{gamma}
\gamma_\alpha = \Gamma_\alpha | 
= 4 \,  \frac{{\cal D}_\alpha {\cal W}_\beta}{{\cal D}^2 {\cal W}^2} \, {\cal W}^\beta \Big{|} 
=  -4 i \left( \frac{{\cal D}_\alpha {\cal W}_\beta}{{\cal D}^2 {\cal W}^2} \Big{|} \right) \eta^\beta \,  , 
\ee 
we will have that 
\be
\gamma_\alpha = \frac{2i {\cal F}_{\alpha \beta}}{{\cal F}^{\rho \sigma} {\cal F}_{\rho \sigma}} \, \eta^\beta + \text{3-fermi terms} \, , 
\ee
where $\eta_\alpha$ and ${\cal F}_{\alpha \beta}$ are defined in \eqref{eta} and \eqref{dw}, respectively. 
Notice that under a local supersymmetry transformation the fermion $\gamma$ transforms as 
\begin{eqnarray}
\delta \gamma_\alpha = - \xi_\alpha(x) + \text{3-fermi terms} \, , 
\end{eqnarray} 
it is in other words a realization of the Volkov--Akulov fermion, 
but in contrast to the latter,  here $\gamma_\alpha$ is  composite.

Now we are ready to reduce the term \eqref{LNEW} to component fields, 
and we find 
\be
\label{component1}
\begin{aligned} 
e^{-1} {\cal L}_\text{NEW} = & - \text{e}^{-2K/3} \, {\cal U}(A^I , \overline A^{\overline J}) 
+ i \, \text{e}^{-2K/3} \, {\cal U} \left( D_m \gamma \, \sigma^m \overline \gamma 
+ D_m \overline \gamma \, \overline \sigma^m \gamma \right) 
\\
& + \text{e}^{-2K/3} \, {\cal U} \lc i \gamma \, \sigma^a \overline \psi_a 
+  \frac23 \overline M \, \gamma^2 
- \frac16 \gamma \sigma^a \overline \gamma \, b_a + {\rm c.c}. 
\rc 
\\
& + \Big{[} \left( \text{e}^{-2K/3} \, {\cal U} \right)_I \lc
\sqrt 2 \, \gamma \chi^I - F^I \, \gamma^2 - i \gamma \, \sigma^m \overline \gamma \, \partial_m A^I 
\rc 
+ {\rm c.c}. \Big{]} 
\\
& + \text{4-fermi terms} \, , 
\end{aligned}
\ee 
where $D_m \gamma_\alpha = \partial_m \gamma_\alpha - \omega_{m \alpha}^{\ \ \ \beta} \gamma_\beta$. 
We see that the first line of \eqref{component1} contains the contribution to the scalar potential and 
also a contribution to the fermion kinetic terms. 
Notice that the second line contains the essential gravitino-goldstino mixing dictated by the Noether method, 
exactly as has been analyzed in \cite{Cribiori:2017laj}. 
Clearly in this setup we can always fix the gauge  
\be
\label{gzero}
\gamma_\alpha = 0 \ \leftrightarrow \ \eta_\alpha = 0 \, , 
\ee
which will eliminate one spin-1/2 fermion from the component form expression and it will be absorbed by the massive gravitino. In this gauge the new Lagrangian term gets a very simple form 
\be
{\cal L}_\text{NEW} \Big{|}_{\eta = 0} = - e \, \text{e}^{-2K/3} \, {\cal U}(A^I , \overline A^{\overline J}) \, . 
\ee
We should stress however that practically imposing the gauge choice \eqref{gzero} 
 requires to solve the equation 
\be
\label{ugauge} 
i \sqrt 2 K_{I \overline J} \overline F^{\overline J} \chi_\alpha^I 
- \frac{i}{\sqrt 2} K_{I \overline L \, \overline J}\, \overline \chi^{\overline L} \overline \chi^{\overline J} \chi_\alpha^I 
+ \sqrt 2 K_{I \overline J} \, \sigma_{\alpha \dot \rho}^a \, \overline \chi^{\dot \rho \overline J} e_a^{\ m} \hat D_m A^I  = 0  ,
\ee
in terms  of a single fermion, say $\chi^1_\alpha$ belonging to a chiral multiplet. The latter is therefore removed from  the spectrum as
in fact it is eaten by the gravitino. 
Finally, 
from the definition of the fermion $\gamma$ in \eqref{gamma} we see that \eqref{component1} is well-defined if and only if \eqref{COND} holds, 
otherwise the fermionic terms become singular.

\subsection{The total supergravity Lagrangian} 

The full theory we are anticipating now, can be written down by coupling 
\eqref{LNEW}  to the standard supergravity \eqref{gen1} so that the dynamics is described by the liberated supergravity  Lagrangian 
\be
\begin{aligned}
\label{LN0}
{\cal L}_{{\rm LIB}} = & {\cal L}_0 + {\cal L}_\text{NEW}  
\\
= & -3 \int {\rm d}^4 \theta \, E \, \text{e}^{-K/3} + 
\left(  \int {\rm d}^2 \Theta \, 2 {\cal E} \, W + {\rm c.c}.  \right) 
\\
& - 16 \, \int {\rm d}^4 \theta \, E \, \text{e}^{-2K/3} \, \frac{
{\cal W}^2(K) \overline {\cal W}^2(K) 
}{{\cal D}^2 {\cal W}^2(K) \overline {\cal D}^2 \overline {\cal W}^2(K)} \ {\cal U}(\Phi^I , \overline \Phi^{\overline J}) \, . 
\end{aligned}
\ee
The theory \eqref{LN0}  describes a four-dimensional ${\cal N}=1$ supergravity coupled to the chiral multiplets $\Phi^I$, 
and the supersymmetry transformations of the component fields are given by \eqref{ddsugra} and \eqref{ddphi}, 
which are the standard local supersymmetry transformations.

Once we integrate out the auxiliary fields, 
and performing the Weyl rescaling of the metric, we find that the bosonic sector is similar to \eqref{BOS}, 
namely 
\be
\label{BOS2} 
e^{-1}  {\cal L} |_\text{bosons}  = - \frac12 R - g_{I \overline J} \p_m A^I \p^m \overline A^{\overline J}  - {\cal V} \, , 
\ee 
but now the scalar potential has the form 
\be
\boxed{
\label{NEWV}
{\cal V} = \text{e}^K \left[ D_{\overline J} \overline W g^{\overline J I} D_I W  - 3 W \overline W \right] 
+ {\cal U}(A^I , \overline A^{\overline J}) 
\, . }
\ee
Clearly if ${\cal U}(A^I , \overline A^{\overline J})$ is positive, 
we have a positive definite uplift, 
and K\"ahler-Weyl invariance is maintained if ${\cal U}$ is inert under K\"ahler transformations. 
The $\text{e}^{-2K/3}$ factor appearing in \eqref{component1} in front of ${\cal U}$ does not appear in 
\eqref{NEWV} because it has been  canceled  by the Weyl rescaling of the vielbein determinant, 
\be
\text{Weyl rescaling}: \quad e \rightarrow \text{e}^{2K/3} \, e \, , 
\ee
needed to write the theory in the Einstein frame as usual.
Notice that  ${\cal L}_\text{NEW}$ changes the on-shell values of the auxiliary fields by 
fermionic corrections compared to their values in standard supergravity. 
Therefore we can now directly evaluate all the terms in ${\cal L}_\text{NEW}$ after eliminating the auxiliary fields. 
Keeping only up to two fermions and after Weyl rescaling we find 
\be
\label{component3}
\boxed{
\begin{aligned} 
e^{-1} {\cal L}_\text{NEW}\Big|_{\rm{ on-shell}} = & - {\cal U}(A^I , \overline A^{\overline J}) 
+ i \, \text{e}^{-K/6} \, {\cal U} \left(  \tilde D_m \tilde \gamma \, \sigma^m \overline{\tilde \gamma} 
+ \tilde D_m \overline{\tilde \gamma} \, \overline \sigma^m \tilde \gamma \right)  
\\
& 
+ {\cal U} \lc 
i \text{e}^{-K/12} \tilde \gamma \, \sigma^a \overline \psi_a 
- 2 \, \overline{W} \text{e}^K \tilde \gamma^2 + {\rm c.c}. 
\rc  
+ \Big{[} {\cal U}_I  \lc
\sqrt 2 \,\text{e}^{-K/12} \tilde \gamma \chi^I 
\right.\\
&\left.
+ \text{e}^{K/3} g^{I \overline J} D_{\overline J} \overline W \, \tilde \gamma^2 
- i \text{e}^{-K/6} \tilde \gamma \, \sigma^m \overline{\tilde \gamma} \, \partial_m A^I 
\rc 
+ {\rm c.c}. \Big{]} 
\\
& 
+ \text{4-fermi terms} \, .  
\end{aligned} 
}
\ee 
Here $\tilde \gamma_\alpha = 2i \tilde {\cal F}_{\alpha \beta} \tilde \eta^\beta / \tilde {\cal F}^{\rho \sigma} \tilde {\cal F}_{\rho \sigma}$ 
with  
\be
\begin{aligned}
 \tilde \eta_\alpha = &- i \sqrt 2 \, \text{e}^{K/4} D_I W \chi_\alpha^I 
+ \sqrt 2 \, \text{e}^{-K/4} g_{I \overline J} \, \sigma_{\alpha \dot \rho}^m \, \overline \chi^{\dot \rho \overline J} \p_m A^I \, , 
\\
\tilde {\cal F}_{\beta \alpha} = & -2 \, \text{e}^{2K/3} \epsilon_{\beta \alpha} D_{\overline J} \overline W g^{\overline J I} D_I W 
+ 2 \, \text{e}^{-K/3} \left( \sigma^n_{\beta \dot \beta} \epsilon^{\dot \beta \dot \alpha} \sigma^m_{\alpha \dot \alpha} \right) 
g_{I \overline J} \,  \partial_m A^I \partial_n \overline A^{\overline J} \, , 
\end{aligned}
\ee
and we have defined the K\"ahler covariant derivative for the composite fermion as 
\be
\tilde D_m \tilde \gamma = D_m \tilde \gamma 
+ \left( \frac16 K_I \partial_m A^I - \frac13 K_{\overline J} \partial_m \overline A^{\overline J} \right) \tilde \gamma \, . 
\ee
The total Lagrangian will be the one describing the standard four-dimensional ${\cal N}=1$ matter-coupled supergravity 
where one simply adds \eqref{component3}. 
To check the K\"ahler-Weyl invariance of \eqref{component3} under \eqref{KSS} one has to take into account 
that the composite fermion $\tilde \gamma$ changes  as
\be
\tilde \gamma \rightarrow \exp \left(-\Sigma + 2 \overline \Sigma \right) \tilde  \gamma \, , 
\ee
while the gravitino and the matter fermions change as usual, namely 
\be
\chi^I \rightarrow \exp \left(3 i \, \text{Im} \Sigma \right) \chi^I   
\, , \quad  
\psi_n \rightarrow \exp \left(-3 i \, \text{Im} \Sigma \right) \psi_n \, . 
\ee
It would be interesting to study these new coupling in a manifestly K\"ahler covariant setup, 
as for example the one presented in \cite{Freedman:2017obq}, 
where the full Lagrangian \eqref{component3} might be easier to write down up to higher order in fermions. 
We leave this interesting calculation for future work. 
Clearly the non-linearities in the fermionic sector which arise while integrating out the auxiliary fields, 
is the price one has to pay for achieving a generic uplift while keeping both  K\"ahler-Weyl invariance of the standard theory, 
and supersymmetry linearly realized off-shell.

Let us note that for the consistent propagation of the gravitino in a curved background one has to ask 
the condition ${\cal V} \geq - 3 |m_{3/2}|^2$ to hold, 
which is always respected by standard supergravity. 
For a discussion of this issue in a supergravity setup see for example \cite{Cribiori:2016qif}. 
Moreover, in standard supergravity this equation is saturated by anti de Sitter supersymmetric vacua, 
where the gravitino supersymmetry transformations \eqref{ddsugra} give the Killing spinors. 
In our setup however, 
the condition ${\cal V} \geq - 3 |m_{3/2}|^2$ 
puts a bound on the value of the function ${\cal U}$ evaluated on a generic background. 
The bound on the background values of ${\cal U}$ is 
\be
\label{UFM}
{\cal U} \geq - \text{e}^K D_{\overline J} \overline W g^{\overline J I} D_I W \, , 
\ee
which is always satisfied for positive definite ${\cal U}$. 
When the bound \eqref{UFM} is saturated on the vacuum, 
we will have an anti de Sitter supergravity, however the gravitino transformations will not provide a full ${\cal N}=1$ Killing spinor, 
unless $\langle D_I W \rangle =0$. 
This is seen either by taking into account that the on-shell $M$ which enters the Killing equations \eqref{ddsugra} is given by \eqref{Mcomp}, 
or simply by the fact that the supersymmetry transformations 
of the matter fermions \eqref{ddphi} will not preserve supersymmetry unless $\langle D_I W \rangle=0$. 
Therefore an ${\cal N}=1$ supersymmetric background will require both $\langle {\cal U} \rangle$ and $\langle D_I W \rangle$ to vanish.

We will now discuss in more detail what happens in the limit 
\be
\label{susyrest}
{\cal F} = \langle g_{I \overline J} F^I \overline F^{\overline J} \rangle \rightarrow 0 \, , 
\ee
where supersymmetry will be restored. 
If we wish to have a generic ${\cal U}$ function entering \eqref{LNEW} then the limit \eqref{susyrest} 
has to be excluded from the moduli space of the effective supergravity theory, 
because all fermionic terms will diverge. 
This signals that when approaching this limit in principle the effective supergravity description will break down. 
However, 
this is not the full story. 
An inspection of the term \eqref{LNEW} will show that a conservative evaluation of the worse possible divergencies 
which can appear in the fermionic sector have the form 
\be
\frac{{\cal U}^{(0)}}{{\cal F}^6} 
\, , \quad \frac{{\cal U}^{(1)}}{{\cal F}^5} 
\, , \quad \frac{{\cal U}^{(2)}}{{\cal F}^4} 
\, , \quad \frac{{\cal U}^{(3)}}{{\cal F}^3} 
\, , \quad \frac{{\cal U}^{(4)}}{{\cal F}^2} \, , 
\ee
where the superscripts ${\cal U}^{(n)}$ refer to derivatives with respect to $A^I$ or $\overline A^{\overline J}$. 
Therefore, 
if we have a function ${\cal U}$ (and its derivatives) which goes to zero faster than $\langle g_{I \overline J} F^I \overline F^{\overline J} \rangle^{n_\text{max}}$, 
then the divergent terms will be damped, 
and the limit where supersymmetry gets restored will exist. 
Thus functions ${\cal U}$ which allow for the restoration of supersymmetry have to satisfy 
\be
\label{limit}
\frac{{\cal U}^{(n)}}{{\cal F}^{6-n}} \Big{|}_{{\cal F} \rightarrow 0} \rightarrow 0 \, . 
\ee 
Under the assumption \eqref{limit} the new terms \eqref{LNEW}  can be always added to a four-dimensional ${\cal N}=1$ supergravity, independent of the properties of the vacuum. 
However, 
if indeed the function scales as in \eqref{limit}, 
then in the supersymmetric point the theory will be identical to standard supergravity, 
as all new interactions will be highly suppressed.

Other deformations of the scalar potential originating from higher derivative couplings are known to exist 
where supersymmetry is generically allowed to be restored \cite{Cecotti:1986jy,Koehn:2012ar,Farakos:2012qu,Koehn:2012np}. 
However our new term has a minimal impact on the bosonic sector of the theory as it changes only the scalar potential 
as shown in \eqref{NEWV}.

\subsection{Gaugings}

As a direct generalization of our construction, 
we can extend the discussion to gauged chiral models. 
However, we will discuss models with gauging only in this subsection and we leave a more detailed discussion for a future work.

The gauging in four-dimensional ${\cal N}=1$ supergravity works by gauging the isometries of the K\"ahler manifold, 
and it is described in detail in standard supergravity textbooks \cite{Wess:1992cp,FvP}. 
Here we follow the approach presented in \cite{Wess:1992cp}. 
In the Lagrangian for the chiral superfields one adds to $K$ a counter-term ${\cal P}$ which renders the theory gauge invariant, 
namely 
\be 
\label{KKP}
K \rightarrow K(\Phi^I, \overline \Phi^{\overline J}) + {\cal P}(\Phi^I, \overline \Phi^{\overline J}, V^{(a)}) \, .  
\ee
The function ${\cal P}$ is uniquely determined by the K\"ahler metric and the isometries one wants to gauge. 
In the Wess--Zumino gauge we have 
\be
{\cal P}(\Phi^I, \overline \Phi^{\overline J}, V^{(a)})  = V^{(a)} \text{D}^{(a)} + \frac12 g_{I \overline J} X^{I(a)} \overline X^{\overline J(b)} V^{(a)} V^{(b)} \, , 
\ee 
where 
\be
\label{dgD}
X^{I(b)} = - i \, g^{\overline J I} \, \partial_{\overline J} \, \text{D}^{(b)}  \, , 
\quad \overline X^{\overline I (b)} = i \, g^{\overline I J} \, \partial_{J} \, \text{D}^{(b)}  \, , 
\ee
are coordinates of the holomorphic (and antiholomorphic) Killing vectors 
\be
X^{(b)} = X^{I(b)} \partial_I \, , \quad  \overline X^{(b)} = \overline X^{\overline I (b)} \partial_{\overline I} \, , 
\ee 
which generate the gauged isometries of the K\"ahler manifold, 
while $\text{D}^{(a)}$ are the Killing potentials. 
The Killing vectors obey the relations 
\be
[X^{(a)},X^{(b)}] = - f^{abc} X^{(c)} \, , \quad [\overline X^{(a)}, \overline X^{(b)}] = - f^{abc} \overline X^{(c)} \, , \quad [X^{(a)}, \overline X^{(b)}] = 0 \, , 
\ee
where $f^{abc}$ are the structure constants of the isometry group. 
The gauge transformations act as 
\be
\begin{aligned}
\delta \Phi^I & = (\Lambda^{(a)} X^{(a)} + \overline \Lambda^{(b)} \overline X^{(b)})  \Phi^I =\Lambda^{(a)} X^{I(a)}(\Phi^J) \, , 
\\
\delta \, \text{e}^{V^{(a)} \text{T}^{(a)}} & =  - i \overline \Lambda^{(a)} \text{T}^{(a)} \, \text{e}^{V^{(a)} \text{T}^{(a)}} 
+ i \, \text{e}^{V^{(a)} \text{T}^{(a)}} \Lambda^{(a)} \text{T}^{(a)} \, . 
\end{aligned} 
\ee
Notice that a gauge transformation does not leave the combination $K+\Gamma$ invariant, 
rather it will generically change it up to a K\"ahler transformation.

In our setup one has to do the same replacement \eqref{KKP} in any place where the K\"ahler potential appears. 
Moreover, we will need to introduce an additional counter-term, 
namely ${\cal Q}(\Phi^I , \overline \Phi^{\overline J}, V^{(a)})$, 
which will render the theory gauge invariant once we set 
\be
{\cal U} \rightarrow {\cal U}(\Phi^I, \overline \Phi^{\overline J}) + {\cal Q}(\Phi^I, \overline \Phi^{\overline J}, V^{(a)}) \, .  
\ee 
To respect all the symmetries of the theory, the sum ${\cal U}+{\cal Q}$ has to be inert under gauge and K\"ahler transformations. 
To achieve this we can impose a series of simple conditions. 
In particular, it has to hold that 
\be
\left( X^{(a)} + \overline X^{(a)}  \right) {\cal U}(\Phi^I , \overline \Phi^{\overline J}) = 0 \, , 
\ee
and under a gauge transformation ${\cal Q}$ has to transform as 
\be
\delta {\cal Q} = i (\Lambda^{(a)} - \overline \Lambda^{(a)}) \tilde{\text{D}}^{(a)}  \, , 
\ee 
where we now define 
\be
\tilde{\text{D}}^{(a)} = i X^{(a)}  {\cal U} = - i \overline X^{(a)}  {\cal U} \, . 
\ee
Following then the general procedure described in \cite{Wess:1992cp} we have in the Wess--Zumino gauge 
\be
{\cal Q} = V^{(a)} \tilde{\text{D}}^{(a)}  + \frac12 {\cal U}_{I \overline J} X^{I(a)} \overline X^{\overline J(b)} V^{(a)} V^{(b)} \, .  
\ee
Notice that now there exists also the equivalent expressions to \eqref{dgD}, 
namely 
\be
{\cal U}_{I \overline J} X^{I(b)} 
= - i \, \partial_{\overline J} \, \tilde{\text{D}}^{(a)}  \, , \quad 
{\cal U}_{J \overline I} \overline X^{\overline I(b)} 
= i \, \partial_{J} \, \tilde{\text{D}}^{(a)}   \, , 
\ee 
and that $\tilde{\text{D}}^{(a)}$ satisfies 
\be
\Big{[}  X^{(a)} + \overline X^{(a)} \Big{]} \tilde{\text{D}}^{(b)}  = - f^{abc} \tilde{\text{D}}^{(c)}  \, , \quad 
X^{(a)} \tilde{\text{D}}^{(b)} + \overline X^{(b)} \tilde{\text{D}}^{(a)}  = 0 \, .  
\ee

We conclude that in a gauged supergravity setup we should consider the term 
\be
\boxed{
\label{LNEWgauged}
{\cal L}_\text{NEW-gauged} = - 16 \, \int {\rm d}^4 \theta \, E \, \text{e}^{-2(K+{\cal P})/3} \, \frac{
{\cal W}^2 \, \overline {\cal W}^2  
}{{\cal D}^2 {\cal W}^2 \, \overline {\cal D}^2 \overline {\cal W}^2} \ ( {\cal U} + {\cal Q} ) \, , 
}
\ee 
where 
\be
{\cal W}_\alpha = - \frac14 \left( \overline{\cal D}^2 - 8 {\cal R} \right) {\cal D}_\alpha (K + {\cal P}) \, . 
\ee 
The bosonic sector of \eqref{LNEWgauged} will simply give 
\be
{\cal L}_\text{NEW-gauged} \Big{|}_\text{bosons} = - e \, \text{e}^{-2K/3} \, {\cal U}(A^I , \overline A^{\overline J}) \, . 
\ee 
Including now \eqref{LNEWgauged} to a gauged chiral model coupled to supergravity, 
the effect on the scalar potential will be exactly given by \eqref{pot-intro}, namely 
\be
\label{gaugedV}
\boxed{
{\cal V} = \frac12 g^2 (\text{D}^{(a)})^2
+ \text{e}^K \left[ D_{\overline J} \overline W g^{\overline J I} D_I W  - 3 W \overline W \right] 
+ {\cal U} \, , }
\ee
where $\cal U$ is now a gauge invariant function. 
Moreover the bosonic sector of such theory will be given by the bosonic sector of standard gauged supergravity, 
except of the scalar potential, which will be given by \eqref{gaugedV}.

Finally the term \eqref{gaugedV} is generically consistent only if  
\be 
\label{condgauge} 
\langle F^I g_{I \overline J} \overline F^{\overline J} \rangle \ne 0 \quad \text{or} \quad  \langle D^{(a)} D^{(a)} \rangle \ne 0 \, , 
\ee 
with $D^{(a)}$ being the auxiliary fields of the vector multiplets. 
The conditions \eqref{condgauge} are essentially the conditions that supersymmetry has to be broken at least from the gauge sector or the matter sector. 
In the limit that supersymmetry gets restored there will be divergencies in the fermionic sector of \eqref{LNEWgauged}. 
However, as we discussed in the previous subsection, it is conceivable that ${\cal U}$ will go to zero fast enough and thus damp 
these divergent terms.

\section{Equivalent formulation with constrained superfields}

Constrained superfields can in principle be used to describe any system where supersymmetry is spontaneously broken. 
A method to find the equivalent theory in terms of constrained superfields has been explained in \cite{Cribiori:2017ngp}. 
We wish to present here the supergravity theory which is equivalent to \eqref{LN0} in terms of constrained superfields. 
We first give the result, 
and then we proceed to do the proof of the equivalence. 
Let us note that for the equivalence to hold we are assuming that supersymmetry is always spontaneously broken, 
which is the generic feature of the models we study here, 
and that ${\cal U} \ne0$.

The Lagrangian \eqref{LN0} is equivalent to a Lagrangian of standard supergravity of the form 
\be
\boxed{
\label{gen22}
{\cal L} =
-3 \int {\rm d}^4 \theta \, E \, \text{e}^{-\hat K/3} + 
\left(  \int {\rm d}^2 \Theta \, 2 {\cal E} \, \hat W + {\rm c.c}.  \right) \, , 
}
\ee
where the new K\"ahler potential and superpotential are given by 
\be
\label{KhatWhat}
\boxed{
\hat K = K + \text{e}^{K} \frac{X \overline X}{{\cal U}}  \, , \quad   
\hat W = W + X \, . 
}
\ee
The chiral multiplets $X$ and $\Phi^I$ in \eqref{gen22} are subject to the following two constraints 
\be
\label{XWXW}
X^2 = 0 \, , \quad   X \, {\cal W}_\alpha(K) = 0 \, . 
\ee 
 The theta expansion  of the chiral superfield $X$ is
\be
X = \frac{(\chi^X)^2}{2 F^X} + \sqrt 2 \, \Theta^\alpha \chi^X_\alpha + \Theta^2 F^X \, ,
\ee
and it scales as $X \rightarrow X \exp\{ - 6 \Sigma\}$ under  the standard K\"ahler-Weyl transformations \eqref{KSS}. 
Notice that the only independent component fields in $X$ are the fermion $\chi^X$ and the auxiliary field $F^X$.

Models for inflation with non-linear realizations of supersymmetry with K\"ahler potentials of the form \eqref{KhatWhat} 
can be found in \cite{McDonough:2016der,Kallosh:2017wnt}. 
Of course, in contrast to \cite{McDonough:2016der,Kallosh:2017wnt}, 
in our setup there is an underlying linear realization of supersymmetry 
which together with K\"ahler-Weyl invariance dictates the form of \eqref{KhatWhat}.

Let us now prove that \eqref{gen22} is equivalent to \eqref{LN0}. 
With a simple manipulation one can bring the Lagrangian \eqref{gen22} to the form 
\be
\label{L0X}
{\cal L} = {\cal L}_0  + {\cal L}_X \, , 
\ee
where ${\cal L}_0$ is the Lagrangian \eqref{gen1} and ${\cal L}_X$ is given by 
\be
\label{LX}
{\cal L}_X = \int {\rm d}^4 \theta E \, \text{e}^{2K/3} \, {\cal U}^{-1}  \, X \, \overline X + 
\left(  \int {\rm d}^2 \Theta \, 2 {\cal E} \, X + {\rm c.c}.  \right) \, . 
\ee
Therefore to prove the equivalence of \eqref{gen22} to \eqref{LN0} we have to reduce \eqref{LX} to \eqref{LNEW}. 
This will indeed happen once we eliminate the auxiliary field $F^X$ of the constrained superfield $X$ via its own equations of motion. 
The variation with respect to $F^X$ can be consistently performed if we split the constrained $X$ multiplet into two independent parts. 
One part which will contain $\chi^X$ as independent component field and one part which will contain $F^X$ as independent component field. 
To this end we split the superfield $X$ as 
\be
\label{split}
X = {\cal Z} \, \overline{\cal H} \, , 
\ee
where ${\cal Z}$ and ${\cal H}$ are chiral superfields ($\overline{\cal D}_{\dot \alpha} {\cal H} = 0 = \overline{\cal D}_{\dot \alpha} X$). 
These superfields are not arbitrary  but rather they are subject to the constraints  
\be
\label{con1}
{\cal Z}^2 = 0 \, , \quad  
- \frac14 {\cal Z} \left( \overline{\cal D}^2 - 8 {\cal R} \right) \overline{\cal Z} = 
{\cal Z} \, , \quad  
{\cal Z} \, {\cal W}_\alpha = 0 \, , 
\ee
and 
\be
\label{XH}
{\cal Z} \, \overline{\cal D}_{\dot \alpha} \overline{\cal H} = 0 \, . 
\ee 
These constraints have been studied in detail in \cite{Ferrara:2016een,DallAgata:2016syy} 
and the splitting \eqref{split} follows the logic discussed in \cite{Cribiori:2017ngp}. 
The superfield ${\cal Z}$ is in fact the one constructed by Lindstr\"om and Ro\v{c}ek in \cite{Lindstrom:1979kq}.

Let us discuss at this point in more detail the splitting \eqref{split}. 
Firstly, we observe that it is consistent if and only if 
\be 
\langle {\cal H}| \rangle \ne 0 \, . 
\ee 
Secondly, the superfield ${\cal Z}$ contains only a single fermion as independent component field, 
which resides in ${\cal D}_\alpha {\cal Z}|$. 
The fermionic degrees of freedom of the  $X$ multiplet, 
namely the fermion $\chi^X$, will now reside in the fermion of ${\cal Z}$. 
Indeed, 
we can see from \eqref{split} that 
\be
{\cal D}_\alpha {\cal Z} | \sim \chi_\alpha^X \, . 
\ee
Thirdly, the constrained chiral superfield ${\cal H}$ contains only a complex scalar, 
namely ${\cal H}|$, as independent component field, 
while the projections ${\cal D}_\alpha {\cal H}|$ and ${\cal D}^2 {\cal H}|$ are composite and do not contain any new independent component fields. 
The auxiliary field of $X$, namely $F^X$, will now reside in the lowest component of ${\cal H}$. 
Indeed we have 
\be
{\cal H}| = \overline{F^X} + \text{fermions}. 
\ee
In this way we can express all the degrees of freedom of $X$ in terms of the ones in ${\cal Z}$ and ${\cal H}$, 
and vice versa. 
Finally, let us point out that the constraint on ${\cal W}_\alpha$ in \eqref{XWXW}, 
namely the constraint $X {\cal W}_\alpha = 0$, now takes the form ${\cal Z} {\cal W}_\alpha = 0$ as one can easily show, 
whereas the constraint $X^2=0$  in \eqref{XWXW} is  reduced to ${\cal Z}^2=0$. 
These constraints are given in \eqref{con1}.

Using \eqref{split} we replace $X$ with ${\cal Z} \, \overline{\cal H}$ in \eqref{LX}. 
After some manipulations the Lagrangian \eqref{LX} can take the form 
\be
\label{inter1}
\begin{aligned}
{\cal L}_X = & \int {\rm d}^4 \theta E \left( \text{e}^{2K/3} \, {\cal U}^{-1}  \, {\cal Z} \, \overline{\cal Z} \, {\cal H} \,  \overline{\cal H} 
+  {\cal Z} \, \overline{\cal Z} \, {\cal H} + 
{\cal Z} \, \overline{\cal Z} \, \overline{\cal H} \right) 
\\
& + \int {\rm d}^4 \theta E \left(G \, {\cal Z} \, \overline{\cal H} + \overline G \, \overline{\cal Z} \, {\cal H}  \right) \, , 
\end{aligned}
\ee
where in the second line of \eqref{inter1} we have introduced the complex linear Lagrange multiplier $G$. 
Being a complex linear superfield, 
$G$ is defined to satisfy 
\be 
\label{D2G} 
(\overline{\cal D}^2 - 8 {\cal R}) G=0 \, . 
\ee
The reason for introducing $G$ in \eqref{inter1} is to make ${\cal H}$ an unconstrained chiral superfield, 
which will allow us to perform the superfield variation and derive its superspace equations of motion. 
If we vary $G$ then we get the constraint \eqref{XH}, and the Lagrangian will be given only by the first line of \eqref{inter1}.

Now we proceed to integrate out $F^X$, 
which amounts to integrating out the chiral superfield ${\cal H}$. 
This is done by performing a superfield variation of ${\cal H}$ and $G$, which gives 
\be
\label{EQH}
\delta {\cal H} \ &:& \ (\overline{\cal D}^2 - 8 {\cal R}) \Big{[} \text{e}^{2K/3} \, {\cal U}^{-1}  \, {\cal Z} \, \overline{\cal Z} \, \overline{\cal H} 
+  {\cal Z} \, \overline{\cal Z} + \overline G \, \overline{\cal Z} \Big{]} = 0 \, , 
\\
\label{EQG}
\delta G \ &:& \ {\cal Z} \, \overline{\cal D}_{\dot \alpha} \overline{\cal H} = 0 \, . 
\ee 
We multiply \eqref{EQH} with $\overline{\cal Z}$ to find 
\be 
\label{ZEQH} 
\text{e}^{2K/3} \, {\cal U}^{-1}  \, {\cal Z} \, \overline{\cal Z} \, \overline{\cal H} 
+  {\cal Z} \, \overline{\cal Z} + \overline G \, \overline{\cal Z} = 0 \, , 
\ee
which we then multiply with ${\cal Z}$ to derive  
\be
\label{ZZG} 
{\cal Z} \, \overline{\cal Z} \, G = 0 \, . 
\ee
Acting on \eqref{ZEQH} with ${\cal Z} \, {\cal D}^2$ and using both the properties \eqref{ZZG} and \eqref{D2G} we find  
\be
\label{dFX}
{\cal Z} \, \overline{\cal Z} \, \overline{\cal H}  = - {\cal Z} \, \overline{\cal Z} \, \text{e}^{-2K/3} \, {\cal U} \, . 
\ee
Equation \eqref{dFX} is the appropriate expression derived from the equations of motion of ${\cal H}$ which 
we can use to eliminate it from the Lagrangian \eqref{inter1}. 
Indeed, inserting \eqref{dFX} into \eqref{inter1} and using also \eqref{EQG}, 
we find 
\be
\label{LNEWZ}
{\cal L}_X = - \int {\rm d}^4 \theta E \, \text{e}^{-2K/3} \, {\cal U}(\Phi^I , \overline \Phi^{\overline J})  \, {\cal Z} \, \overline{\cal Z} \, . 
\ee
We remind the reader that $\cal Z$ in \eqref{LNEWZ} is a constrained chiral superfield which satisfies \eqref{con1}.

To complete the equivalence we have to replace ${\cal Z}$ in \eqref{LNEWZ} 
with an expression in terms of ${\cal W}_\alpha$, 
such that \eqref{LNEWZ} takes the form \eqref{LNEW}. 
As we said, $\cal Z$ satisfies the constraints \eqref{con1} which we can use to express ${\cal Z}$ in terms of ${\cal W}_\alpha$. 
Acting with ${\cal D}^2$ on the third constraint of \eqref{con1}, 
we get that 
\be
\label{WZZ}
{\cal W}_\alpha(K) = \frac{-2 {\cal D}^\beta {\cal Z} {\cal D}_\beta {\cal W}_\alpha(K) - {\cal Z} {\cal D}^2 {\cal W}_\alpha(K)}{{\cal D}^2 {\cal Z}} \, , 
\ee
which leads to 
\be
\label{ZZWW}
{\cal Z} \overline{\cal Z} = 16 \frac{{\cal W}^2 \overline{\cal W}^2}{{\cal D}^2{\cal W}^2 \overline{\cal D}^2\overline{\cal W}^2} \, . 
\ee
Replacing \eqref{ZZWW} into \eqref{LNEWZ}  we reproduce  \eqref{LNEW}.  
In other words,  the Lagrangian \eqref{L0X} is equivalent to \eqref{LN0},  
 and therefore \eqref{gen22} is equivalent to \eqref{LN0} as well.

\section{A single chiral superfield}

In this section we study the coupling of a single chiral superfield $\Phi$ to supergravity. 
We will assume that the dynamics is described by the Lagrangian  \eqref{LN0} 
and we will study the properties of this model in two steps. 
First we study the model in the unitary gauge where it simplifies considerably. 
Second we study the superspace equations derived from the variations of the auxiliary fields and we see that they have a very interesting property. 
Namely, 
the on-shell value of the auxiliary fields is determined only by their on-shell values derived from Lagrangian \eqref{gen1}, 
and from consistency conditions derived from the properties of the supergravity algebra.

The data for a single chiral multiplet $\Phi$ coupled to supergravity described by the Lagrangian \eqref{LN0} are the K\"ahler potential, the superpotential and the uplifting potential $K,W$ and $\mathcal{U}$, respectively, which are functions of  $\Phi$ 
\be
K = K(\Phi, \overline \Phi) \, , \quad  W = W(\Phi) \, , \quad  {\cal U} = {\cal U}(\Phi , \overline \Phi) \, . 
\ee 
To simplify formulas in component form we use the definitions for the component fields 
of the single chiral superfield given by: $\Phi = A + \sqrt 2 \Theta \chi + \Theta^2 F$. 
In the case of a single chiral multiplet the equation \eqref{ugauge}
can easily be solved and leads to
\be
\eta_\alpha = 0 \  \rightarrow \  \chi_\alpha = 0 .
\ee
Therefore, in the unitary gauge, the full supergravity Lagrangian in component form including the uplifting term      is 
\be 
\label{EX}
\begin{aligned} 
e^{-1} {\cal L} = & - \frac12 R - K_{A \overline A} \p_m A \p^m \overline A  - {\cal V}(A , \overline A) 
+ \frac14 \epsilon^{klmn}  ( K_A \partial_k A - K_{\overline A} \partial_k \overline A ) \, \psi_l \sigma_m \overline \psi_n 
\\
& + \frac{1}{2} \epsilon^{klmn} (\overline \psi_k \overline \sigma_l D_m \psi_n - \psi_k \sigma_l D_m \overline \psi_n) 
- \text{e}^{K/2} W \, \overline \psi_a \overline \sigma^{ab} \overline \psi_b 
- \text{e}^{K/2} \overline W \, \psi_a  \sigma^{ab}  \psi_b \, , 
\end{aligned}
\ee
where 
\be
{\cal V}(A , \overline A) =  \text{e}^K \left[ D_{\overline A} \overline W (K_{A \overline A})^{-1} D_A W  - 3 W \overline W \right] + {\cal U}(A , \overline A) \, . 
\ee
For the consistency of this theory however we must have
\be
\label{DW}
\langle D_A W \rangle \ne 0 \, . 
\ee
This is of course not manifest in \eqref{EX}, 
but one has to keep in mind that to write \eqref{EX} we have already assumed \eqref{DW} holds so that we can perform all the redefinitions and eliminate the goldstino. 
To study the limit where $\langle D_A W \rangle \rightarrow 0$ one has to go out of the gauge $\eta=0$, 
where clearly the theory becomes ill-defined for $\langle D_A W \rangle \rightarrow 0$.

Let us note that if we give to the K\"ahler potential and the superpotential the no-scale 
form \cite{Cremmer:1983bf}, namely $K = -3 \, \text{ln} (T + \overline T)$ and  $W = W_0$, 
then the scalar potential simplifies to ${\cal V} = {\cal U}$. 
In contrast, if we set ${\cal U} = - \text{e}^K \left[ D_{\overline A} \overline W (K_{A \overline A})^{-1} D_A W  - 3 W \overline W \right]$, 
then we will have ${\cal V}=0$ independent of $K$ and $W$. 
Finally, if we have ${\cal U} =3 \text{e}^K W \overline W$ the scalar potential 
becomes ${\cal V} = \text{e}^K D_{\overline A} \overline W (K_{A \overline A})^{-1} D_A W$.

\subsection{The properties of the auxiliary fields}

The above setup  for a single chiral superfield has a tight structure, 
which as we already mentioned has, among others, a very interesting property concerning the auxiliary fields. 
As we will see, the new couplings we have introduced change the on-shell values of the auxiliary fields in a very constrained way. 
In particular, 
we will first present a set of constraints which, for the case of a single chiral superfield, 
are trivially satisfied by the standard supergravity on-shell auxiliary fields. 
Then we will show that the on-shell values of the liberated supergravity auxiliary fields satisfy exactly the same constraints, 
which are generically solved iteratively. 
As a result, 
the on-shell values of the liberated supergravity auxiliary fields are uniquely derived by their on-shell values 
in standard supergravity via a straightforward iterative procedure. 
This is a profound property which means that a similar setup could be also possible in supergravity theories where the off-shell structure is not yet complete. 
This is one of the most promising future directions of our work.

Let us now see how the on-shell properties of the auxiliary fields are derived. 
In standard supergravity (${\cal U} \equiv 0$) the superspace equations of motion for the 
supergravity and matter superfields read\footnote{These equations can be derived with superspace methods which we will discuss in the next subsection. Alternatively one can vary the auxiliary fields of each multiplet and lift the equations from component fields to full superspace. 
Both methods give essentially the same results.}  
\be
\label{FEQ}
W_\Phi -\frac14 \left( \overline{\cal D}^2 - 8 {\cal R} \right) \lc \text{e}^{-K/3} K_\Phi \rc & = & 0\, , 
\\
\label{MEQ}
W + \frac14 \left( \overline{\cal D}^2 - 8 {\cal R} \right) \lc \text{e}^{-K/3} \rc  & = & 0\, , 
\\
\label{bEQ}
{\cal G}_{\alpha \dot \alpha} - \frac{1}{4} \text{e}^{K/3} \left( \ls {\cal D}_\alpha , \overline{\cal D}_{\dot \alpha} \rs \lc \text{e}^{-K/3} \rc 
- 3 \lc \text{e}^{-K/3} \rc_{\Phi \overline \Phi} {\cal D}_\alpha \Phi \overline{\cal D}_{\dot \alpha} \overline \Phi 
\right)& = & 0\, . 
\ee
Notice the two first equations are chiral whereas the third is real. 
The auxiliary fields of the supergravity and the matter multiplet are given by 
\be
F = -\frac14 {\cal D}^2 \Phi | \, , \quad   M = -6 \, {\cal R} | \, , \quad   b_a = -3 \, {\cal G}_a | \, , 
\ee
therefore their on-shell values are essentially determined by the lowest components of the superspace 
equations \eqref{FEQ}, \eqref{MEQ} and \eqref{bEQ} respectively. 
In particular \eqref{FEQ} can be recast in the form 
\be
\begin{aligned}
\label{Frecast} 
(K_\Phi K_{\overline \Phi} - 3 K_{\Phi \overline \Phi}  ) \, \overline{\cal D}^2  \overline \Phi = & 
- 12 \lc W_\Phi \text{e}^{K/3} + 2 {\cal R} K_\Phi \rc 
\\
& + \lc \frac13 K_\Phi (K_{\overline \Phi})^2 
- K_\Phi K_{\overline \Phi \, \overline \Phi} 
- 2 K_{\overline \Phi} K_{\Phi \overline \Phi} 
+ 3 K_{\Phi \overline \Phi \, \overline \Phi} 
\rc ({\overline{\cal D} \, \overline \Phi})^2 \, , 
\end{aligned}
\ee
which gives the equations for the auxiliary field $F$ when we project to componentns. 
Equivalently, when we act with ${\cal D}_\alpha$ on \eqref{FEQ} or \eqref{Frecast}  
and project to components we get the equations of motion of the fermion $\chi_\alpha$. 
If we act with ${\cal D}^2$ we get the equations of motion for the complex scalar $A$. 
Equation \eqref{MEQ} can be recast in the form 
\be
\label{Mrecast} 
{\cal R} = \frac12 W \text{e}^{K/3} + \frac18  \text{e}^{K/3} \overline{\cal D}^2 \text{e}^{-K/3}  \, , 
\ee
which gives the equation for $M$ once we project to components.\footnote{In standard supergravity it is preferable to also use 
equation \eqref{Mrecast} to bring \eqref{Frecast} to a simpler form.} 
Finally, 
equation \eqref{bEQ} clearly gives the equations for $b_a$ when projected to components. 
The ${\cal D}_\alpha$ component of \eqref{bEQ} will give the gravitino equations of motion, 
while the $[{\cal D}_\alpha , \overline{\cal D}_{\dot \alpha}]$ component gives the Einstein equations. 
Further properties of equation \eqref{bEQ} in component form can be found for example in \cite{Ferrara:2017yhz}.

However when supersymmetry is spontaneously broken, 
i.e. $\langle F \rangle \ne 0$, 
we can define the superfield $\Gamma_\alpha$ as in \eqref{WL}, 
and we can also construct the nilpotent chiral superfield ${\cal Z}$ of \cite{Lindstrom:1979kq} as follows 
\be
\label{ZGG}
{\cal Z} = - \frac14 \left( \overline{\cal D}^2 - 8 {\cal R} \right) \Gamma^2 \overline \Gamma^2 \, . 
\ee
Then, we simply multiply the equations \eqref{FEQ} and \eqref{MEQ} with ${\cal Z}$, 
to derive 
\be
\boxed{
\label{FEQG} 
{\cal Z} \, \Big{[} W_\Phi -\frac14 \left( \overline{\cal D}^2 - 8 {\cal R} \right) \lc \text{e}^{-K/3} K_\Phi \rc \Big{]} =  0 \, 
}
\ee
and 
\be
\boxed{
\label{MEQG}
{\cal Z} \, \Big{[} W + \frac14 \left( \overline{\cal D}^2 - 8 {\cal R} \right) \lc \text{e}^{-K/3} \rc \Big{]} =  0 \, , 
}
\ee
respectively. 
Finally, multiplying  \eqref{bEQ}  with ${\cal Z} \overline{\cal Z}$ we obtain
\be
\boxed{
\label{bEQG}
{\cal Z} \overline{\cal Z} \, \Big{[} {\cal G}_{\alpha \dot \alpha} - \frac{1}{4} \text{e}^{K/3} \left( \ls {\cal D}_\alpha , \overline{\cal D}_{\dot \alpha} \rs \lc \text{e}^{-K/3} \rc 
- 3 \lc \text{e}^{-K/3} \rc_{\Phi \overline \Phi} {\cal D}_\alpha \Phi \overline{\cal D}_{\dot \alpha} \overline \Phi 
\right)
\Big{]} =  0\, . 
}
\ee
Equations \eqref{FEQG}-\eqref{bEQG} have well understood properties which have been explained in \cite{DallAgata:2016syy} for a generic setup, 
and for the supergravity auxiliary fields in particular in \cite{Cribiori:2016qif,Ferrara:2017bnq}. 
Their effect is to completely eliminate the auxiliary fields from the spectrum, 
by giving them the values determined uniquely by solving \eqref{FEQG}-\eqref{bEQG} iteratively. 
Therefore, 
even though these equations are completely compatible and derivable from standard supergravity, 
they can be viewed as independent equations which determine the on-shell values of the auxiliary fields $F$, $M$ and $b_a$.

The important result now is that the superspace equations derived from \eqref{LN0}, 
for a single chiral superfield, 
reproduce exactly \eqref{FEQG}-\eqref{bEQG}. 
In other words the variations arising from \eqref{LNEW} 
have the profound property to leave the equations \eqref{FEQG}-\eqref{bEQG} unchanged. 
We will prove this in the next subsection. 
For the rest of this subsection, 
we will illustrate the properties of these equations.

To explain the structure of equations \eqref{FEQG}-\eqref{bEQG}, 
and their relation to \eqref{FEQ}-\eqref{bEQ} in standard supergravity, 
let us use \eqref{FEQG} as an example which is also familiar to most readers. 
To avoid long formulas, let us define 
\be
{\cal Y} = W_\Phi -\frac14 \left( \overline{\cal D}^2 - 8 {\cal R} \right) \lc \text{e}^{-K/3} K_\Phi \rc \, . 
\ee
Here ${\cal Y}$ is a composite chiral superfield, 
with component fields  
\be
{\cal Y} = Y + \sqrt 2 \Theta \chi^Y + \Theta^2 F^Y \, . 
\ee
As we explained earlier, 
the superfield ${\cal Y}$ has the property to contain the matter multiplet equations of motion in its three component fields, 
namely 
\be
\begin{aligned}
Y = 0 \  &\rightarrow& \text{Variational equation of $F$} \, , 
\\
\chi^Y = 0 \  &\rightarrow& \text{Equations of motion of $\chi$} \, , 
\\
F^Y = 0 \  &\rightarrow& \text{Equations of motion of $A$} \, . 
\end{aligned}
\ee
The constraint \eqref{FEQG} will now give 
\be
\label{ZY}
{\cal Z} \, {\cal Y} = 0 \, , 
\ee
which is solved by 
\be
\label{XYsol}
Y = \frac{\chi^{\cal Z} \chi^{Y}}{F^{\cal Z}} - \frac{\cal Z}{F^{\cal Z}} F^Y \, .   
\ee 
In standard supergravity, 
clearly one finds $Y=0$ once the equations of motion for the physical fields $\chi$ and $A$ are 
assumed.\footnote{This is related to the nomenclature ``on-shell'', used to refer to the supergravity theories where the auxiliary fields have been integrated out.}  
The equation $Y=0$ will just set the auxiliary field to its form derived by standard supergravity.

If supersymmetry is unbroken, 
equation \eqref{ZY} is satisfied trivially for $Y=0$, $\chi^Y=0$ and $F^Y=0$. 
However when supersymmetry is spontaneously broken, 
equation \eqref{ZY} can be still satisfied off-shell if $Y$ is given by the condition \eqref{XYsol}. 
In particular \eqref{ZY} will lead to a superspace equation of the form 
\be
\begin{aligned}
\label{FrecastG} 
- \frac14 \overline{\cal D}^2  \overline \Phi = & \ 
3  (K_\Phi K_{\overline \Phi} - 3 K_{\Phi \overline \Phi}  )^{-1} \! 
\lc W_\Phi \text{e}^{K/3} + 2 {\cal R} K_\Phi \rc 
\\
& - \frac14 (K_\Phi K_{\overline \Phi} - 3 K_{\Phi \overline \Phi}  )^{-1} \! 
\lc \frac13 K_\Phi (K_{\overline \Phi})^2 
- K_\Phi K_{\overline \Phi \, \overline \Phi} 
- 2 K_{\overline \Phi} K_{\Phi \overline \Phi} 
+ 3 K_{\Phi \overline \Phi \,  \overline \Phi} 
\rc \! ({\overline{\cal D} \, \overline \Phi})^2 
\\ 
& + {\cal O} \left( \Gamma, \overline \Gamma \right) \, . 
\end{aligned}
\ee 
In the above equation we have denoted by  ${\cal O} \left( \Gamma, \overline \Gamma \right)$ terms which 
contain at least one $\Gamma_\alpha$ or $\overline \Gamma_{\dot \alpha}$ superfield, and are uniquely determined by \eqref{ZY}. 
In this case the equation for the auxiliary field $F$ will be given by the lowest component field projection of \eqref{FrecastG}, 
which leads to an expression of the form 
\be
\boxed{
\label{FFGG} 
F  = F_0
+ {\cal O} \left( \gamma, \overline \gamma \right) \, . 
}
\ee
Here  $F_0$ refers to the one-shell value of the auxiliary field $F$ when the chiral multiplet is coupled 
to standard supergravity (the one determined by \eqref{Frecast}). 
In addition, 
we have denoted by  ${\cal O} \left( \gamma, \overline \gamma \right)$ the component projection 
of the ${\cal O} \left( \Gamma, \overline \Gamma \right)$ terms in \eqref{FrecastG}. 
Of course the ${\cal O} \left( \gamma, \overline \gamma \right)$ terms will contain all the component fields of the theory (including $F$), 
and therefore \eqref{FFGG} has to be solved iteratively. 
Once this is done we will have: 
$F |_\text{on-shell}  = F_0
+ {\cal O} \left( \gamma, \overline \gamma \right),$ 
and therefore on-shell $F$ will be given as a function of the remaining component fields of the theory.

To summarize, 
equation \eqref{FFGG} is controlled only from the structure of the supergravity algebra and 
the form of $F_0 $. 
A similar discussion can be done also for equations \eqref{MEQG} and \eqref{bEQG}, 
giving 
\be
\label{MMGG}
\boxed{
M  = M_0
+ {\cal O} \left( \gamma, \overline \gamma \right) 
}
\ee
and 
\be
\label{bbGG}
\boxed{
b_a  = b_{a \, 0} 
+ {\cal O} \left( \gamma, \overline \gamma \right) 
}
\ee
for the on-shell values of the supergravity auxiliary fields. 
Here $M_0$ and $b_{a \, 0}$ refer to their values in standard matter-coupled supergravity. 
As a result, 
one can solve equations \eqref{FFGG}, \eqref{MMGG} and \eqref{bbGG} iteratively. 
The only input is then the solutions for the auxiliary fields derived from the standard matter-coupled supergravity Lagrangian 
and the structure of the supergravity algebra which controls the iterative formulas.\footnote{It would be interesting to investigate the properties of these equations within the setup presented in \cite{Vanhecke:2017chr}.}

\subsection{Deriving the equations for the auxiliary fields}

Let us now prove that   supergravity  coupled to a single chiral superfield $\Phi$ in the presence of \eqref{LNEW} 
leads to \eqref{FEQG}-\eqref{bEQG}. 
To prove this we have to perform three independent superfield variations 
and multiply the resulting equation with ${\cal Z}$ or ${\cal Z} \overline{\cal Z}$. 
Then \eqref{FEQG}-\eqref{bEQG} hold if all contributions from  \eqref{LNEW} vanish.

\subsubsection*{The variations for the auxiliary fields $F$ and $M$:}

From the construction \eqref{ZGG} we have the identities  
\be
&& {\cal Z} \, {\cal W}_\alpha(K) = 0 \, , 
\\
\label{ZGW}
&& {\cal Z} \overline{\cal Z} = \Gamma^2 \overline \Gamma^2 
= 16 \frac{{\cal W}^2 \overline{\cal W}^2}{{\cal D}^2{\cal W}^2 \overline{\cal D}^2\overline{\cal W}^2} \, , 
\\
&& {\cal Z} \overline{\cal Z} \,  {\cal D}_\alpha \Phi = 0 = {\cal W}^2 \overline{\cal W}^2 {\cal D}_\alpha \Phi  \, , 
\ee
which we will use throughout the calculations that follow.

Let us start from the variation of the full superspace Lagrangian of liberated supergravity \eqref{LN0}, 
which will give the superspace equations relevant to $F$. 
This is achieved in a superspace setup by varying the chiral superfield $\Phi$ by using the 
form $\delta \Phi = ( \overline{\cal D}^2 - 8 {\cal R} ) \delta N$, 
for some complex unconstrained prepotential $N$. 
The result of the variation corresponding to $F$ is 
\be
\begin{aligned}
&
W_\Phi -\frac14 \left( \overline{\cal D}^2 - 8 {\cal R} \right) \lc \text{e}^{-K/3} K_\Phi \rc 
+ 4 \left( \overline{\cal D}^2 -8 {\cal R} \right) 
\Big{[}   
 (\text{e}^{-2K/3}{\cal U} )_\Phi \, \frac{
{\cal W}^2 \overline {\cal W}^2 
}{{\cal D}^2 {\cal W}^2 \overline {\cal D}^2 \overline {\cal W}^2} 
\Big{]}
\\
& -2  (\overline{\cal D}^2 -8 {\cal R}) \left\{ K_\Phi 
{\cal D}_\alpha 
\left( \overline{\cal D}^2 -8 {\cal R} \right) 
\Big{[} {\cal W}^\alpha 
\frac{\overline {\cal W}^2  (\text{e}^{-2K/3}{\cal U} )}{|{\cal D}^2 {\cal W}^2 |^2} 
- {\cal W}^\alpha  {\cal D}^2 \lc 
\frac{\overline {\cal W}^2  {\cal W}^2 (\text{e}^{-2K/3}{\cal U} )}{|{\cal D}^2 {\cal W}^2 |^2 {\cal D}^2 {\cal W}^2 } 
\rc 
\Big{]} \right. 
\\
& \left.
\qquad 
\qquad 
\qquad 
\quad 
+ K_{\Phi \overline \Phi} (\overline{\cal D}_{\dot \rho} \overline \Phi) 
({\cal D}^2 - 8 {\cal R}) 
\left( {\cal W}^2 
\overline{\cal W}^{\dot \rho} 
\Big{[}
\frac{\text{e}^{-2K/3}{\cal U}}{|{\cal D}^2 {\cal W}^2 |^2} 
- \overline{\cal D}^2 \lc 
\frac{\overline{\cal W}^2 \text{e}^{-2K/3}{\cal U} }{|{\cal D}^2 {\cal W}^2 |^2 \overline{\cal D}^2 \overline{\cal W}^2 } 
\rc 
\Big{]} \right) \right\} = 0 \, . 
\end{aligned}
\ee 
Multiplying the above equation by ${\cal Z}$ and using the identities 
\be
\begin{aligned}
& {\cal Z} \left( \overline{\cal D}^2 -8 {\cal R} \right) 
\Big{[}   
 (\text{e}^{-2K/3}{\cal U} )_\Phi \, \frac{
{\cal W}^2 \overline {\cal W}^2 
}{{\cal D}^2 {\cal W}^2 \overline {\cal D}^2 \overline {\cal W}^2} 
\Big{]} \equiv 0 \, , 
\\
& {\cal Z} \ 
(\overline{\cal D}^2 -8 {\cal R})K_\Phi  \ 
{\cal D}^\alpha {\cal W}_\alpha \ 
\left( \overline{\cal D}^2 -8 {\cal R} \right) 
\Big{[}
\frac{\overline {\cal W}^2  (\text{e}^{-2K/3}{\cal U} )}{|{\cal D}^2 {\cal W}^2 |^2} 
- {\cal D}^2 \lc 
\frac{\overline {\cal W}^2  {\cal W}^2 (\text{e}^{-2K/3}{\cal U} )}{|{\cal D}^2 {\cal W}^2 |^2 {\cal D}^2 {\cal W}^2 } 
\rc 
\Big{]} \equiv 0 \, , 
\\
& {\cal Z} \ (\overline{\cal D}^2 -8 {\cal R}) K_{\Phi \overline \Phi} \ \overline{\cal D}_{\dot \rho} \overline \Phi \ 
{\cal D}^2 {\cal W}^2 \, 
\overline{\cal W}^{\dot \rho} 
\Big{[}
\frac{\text{e}^{-2K/3}{\cal U}}{|{\cal D}^2 {\cal W}^2 |^2} 
- \overline{\cal D}^2 \lc 
\frac{\overline{\cal W}^2 \text{e}^{-2K/3}{\cal U} }{|{\cal D}^2 {\cal W}^2 |^2 \overline{\cal D}^2 \overline{\cal W}^2 } 
\rc 
\Big{]} \equiv 0 \, , 
\end{aligned}
\ee
we get  \eqref{FEQG} which is the constraint obeyed by standard supergravity.

Now we turn to the variation of the auxiliary field $M$. 
We want to show that all the contributions to the variation of $\overline M$ arising from \eqref{LNEW} 
are vanishing once we multiply with ${\cal Z}$. 
The easiest way to derive the variational equations of $M$ is to turn to the super-Weyl formulation of supergravity and vary with respect to the compensator. 
Assume we have the chiral superfield $\sigma$ as compensator, 
which transforms under the super-Weyl transformation \eqref{SWI} as 
\be
\sigma \rightarrow \sigma + \Sigma  
\ee
and under K\"ahler as: $\sigma \rightarrow \sigma - \Sigma$. 
The Lagrangian \eqref{LNEW} then takes the super-Weyl invariant form 
\be
\label{LNEWs}
{\cal L}_\text{NEW} = - 16 \, \int {\rm d}^4 \theta \, E \, \text{e}^{-4 \sigma -4 \overline\sigma} \, \text{e}^{-2K/3} \, \frac{
{\cal W}^2(K) \overline {\cal W}^2(K) 
}{{\cal D}^2 {\cal W}^2(K) \overline {\cal D}^2 \overline {\cal W}^2(K)} \ {\cal U}(\Phi^I , \overline \Phi^{\overline J})  \, . 
\ee
To go back to standard supergravity we gauge-fix $\sigma =0$. 
We now vary the full Weyl-invariant liberated supergravity Lagrangian with respect to $\sigma$ 
(and then we set $\sigma =0$) to get 
\be
\label{fullM}
W + \frac14 \left( \overline{\cal D}^2 - 8 {\cal R} \right) \lc \text{e}^{-K/3} \rc 
- \frac83 \, 
\left( \overline{\cal D}^2 -8 {\cal R} \right) 
\Big{[}  \text{e}^{-2K/3} \, \frac{
{\cal W}^2  \overline {\cal W}^2  
}{{\cal D}^2 {\cal W}^2 \overline {\cal D}^2 \overline {\cal W}^2} \ {\cal U} 
\Big{]} = 0 \, . 
\ee 
Once we multiply with ${\cal Z}$ and using the identity  
\be
{\cal Z}  
\left( \overline{\cal D}^2 -8 {\cal R} \right) 
\Big{[}  \text{e}^{-2K/3} \, \frac{
{\cal W}^2  \overline {\cal W}^2  
}{{\cal D}^2 {\cal W}^2 \overline {\cal D}^2 \overline {\cal W}^2} \ {\cal U} 
\Big{]} 
\equiv 0 \, , 
\ee
we find \eqref{MEQG}. 
Note that equation \eqref{fullM} is related to the component field variation of $\overline M$.

\subsubsection*{The variation for the auxiliary field $b_a$:}

Finally, we have to perform the variation of the $b_a$ in Lagrangian \eqref{LN0} and show that it does not alter \eqref{bEQG}. 
In principle one can follow the procedure presented in \cite{Buchbinder:1998qv,Bandos:2011fw} and perform a full superspace variation to derive the equivalent equations for the ${\cal G}_a$ superfield.

However, 
since in our setup supersymmetry will be generically spontaneously broken, 
we can utilize the form of the action we derived in the previous section in terms of constrained superfields. 
Therefore instead of varying $b_a$ in \eqref{LN0}, 
we can equivalently vary $b_a$ in 
\be
\label{D7}
{\cal L} = {\cal L}_0 +  {\cal L}_{X} +  {\cal L}_\text{L} \, , 
\ee
where the term ${\cal L}_\text{L}$ contains the appropriate chiral Lagrange multipliers $\rho$ and $\tau^\alpha$, 
namely 
\be
\label{LD7}
{\cal L}_\text{L} = \left(  \int {\rm d}^2 \Theta \, 2 {\cal E} \, \rho \, X^2  + {\rm c.c}.  \right)  
+ \int {\rm d}^4 \theta E \left(X \, \tau^\alpha \, {\cal D}_\alpha K + {\rm c.c}. \right) \, . 
\ee 
In the Lagrangian \eqref{D7} the superfield $X$ is chiral but otherwise unconstrained, 
and the same holds for $\Phi$. 
However once we vary $\rho$ and $\tau$ we get the equations \eqref{XWXW}. 
Since now we have only standard chiral superfields in the theory we easily reduce to components and perform the standard supergravity 
variation which will give the equations for $b_a$. 
In particular the only $b_a$ dependence of \eqref{D7} is given by the following terms 
\be
\begin{aligned}
{\cal L}_\text{L} =  & \lc \frac{\p }{\p \Theta^\beta} \left( {\cal E} \,  X \, \tau_\beta \right) \Big{|} \rc 
\, K_{A \overline A} \, b_a \, \chi \s^a \overline \chi \, + {\rm c.c}. 
\\
& - \lc  {\cal E} \,  X \, \tau^\alpha \Big{|} \rc \, \Big{[} 
i K_{A \overline A} \, \s^c_{\alpha \dot \beta} e_c^m \, \overline \psi_m^{\dot \beta} \, b_a \, \chi \s^a \overline \chi 
- b_{\alpha \dot \beta} \left(  \sqrt 2 i \, K_{A \overline A} \,F \, \overline \chi^{\dot \beta} 
- \frac{i}{\sqrt 2} K_{\overline A A A} \chi^2 \overline \chi^{\dot \beta} \right. 
\\ 
& \qquad  \qquad \qquad  \qquad + \sqrt 2 \, K_{A \overline A} \, \overline \s^{a \dot \beta \rho} \chi_\rho \, \hat D_a \overline A  
\Big{]} + {\rm c.c}. + \text{ terms with no $b_a$ dependence} \, . 
\end{aligned}
\ee 
It is now straightforward to show that the variation of the Lagrangian \eqref{D7} 
under $b^a$ 
will give component field equations which can be directly lifted to the superspace equations of the form 
\be
\label{dH}
{\cal G}_{\alpha \dot \alpha} - \frac{1}{4} \text{e}^{K/3} \left( \ls {\cal D}_\alpha , \overline{\cal D}_{\dot \alpha} \rs \lc \text{e}^{-K/3} \rc 
- 3 \lc \text{e}^{-K/3} \rc_{\Phi \overline \Phi} {\cal D}_\alpha \Phi \overline{\cal D}_{\dot \alpha} \overline \Phi 
\right) 
= {\cal D}^\beta X \, {\cal B}_{\beta \alpha \dot \alpha} + {\rm c.c}. 
\ee
for some superfield ${\cal B}_{\beta \alpha \dot \alpha}$. 
From \eqref{XWXW} however we have 
\be
{\cal W}^2 \overline{\cal W}^2 \, {\cal D}_\alpha X = 0 \ \rightarrow \  {\cal Z} \overline{\cal Z} \, {\cal D}_\alpha X = 0 \, , 
\ee
where we have used \eqref{ZGW}. 
Then we can derive \eqref{bEQG} once we multiply \eqref{dH} with ${\cal Z} \overline{\cal Z}$. 
We therefore conclude that the variation of \eqref{LN0} with respect to $b_a$ gives \eqref{bEQG}. 
Notice that even though \eqref{dH} will depend on ${\cal U}$ (through ${\cal B}_{\beta \alpha \dot \alpha}$), 
equation \eqref{bEQG} does not.

\section{Discussion and outlook}

In this work we have studied a new deformation of four-dimensional ${\cal N}=1$ matter-coupled 
supergravity which has the effect of directly adding an arbitrary real 
function ${\cal U}(A^I , \overline A^{\overline J})$ to the scalar potential. 
Our proposal works for any generic gauged chiral model and it is valid under any circumstance 
as long as supersymmetry is spontaneously broken in some sector. 
The superspace term preserves the K\"ahler-Weyl invariance of the standard supergravity theory 
and supersymmetry is linearly realized off-shell. 

Summarizing, the features of such liberated ${\cal N}=1$ supergravity we discussed here comprise the following unique features: 
\begin{itemize}
\item It is invariant under $\mathcal{N}=1$ linear local supersymmetry off-shell. 
\item It is generically in a supersymmetry broken phase. 
\item It preserves the K\"ahler-Weyl invariance of standard 
$\mathcal{N}=1$ supergravity.
\item  The potential of the theory has the standard structure of the $\mathcal{N}=1$ theory with an additional uplifting part.
\item Depending of the behavior of $\cal{U}$ as $ F \to 0$, we have two cases. Either the new term does not vanish and  we are always in the broken phase, or it vanishes giving standard $\mathcal{N}=1$ supergravity with higher order interaction terms. In particular, in the latter case, its vacuum structure is the same as in the standard $\mathcal{N}=1$ case.   
\end{itemize}

As mentioned in the introduction, liberated supergravity is rich of consequences. For instance, it is well known that inflationary models rooted in supergravity suffer from the so-called $\eta$ problem \cite{LR}. 
The latter arises because the flatness of the standard scalar potential is easily ruined by the overall exponential term containing the K\"ahler potential. 
Since the new extra term \eqref{LNEW} is completely free from constraints one has the freedom to write any potential which could support an inflationary dynamics. Furthermore, liberated supergravity is a suitable starting point to construct low-energy phenomenological models where the fermion-boson mass degeneracy is broken by the extra terms. 
This represents an alternative (or extension) to the traditional low-energy supergravity construction where 
the degeneracy is broken by soft terms. 

The new term presented here and the term in \cite{Cribiori:2017laj}  
are only the first and probably simplest constructions which one can encounter, 
but there are definitely more possibilities even in the minimal four-dimensional ${\cal N}=1$ supergravity. 
Moreover there has to be generalizations with more supersymmetry and/or in other dimensions; 
in other words we have been only scratching the surface here. 
However, 
the fact that the function ${\cal U} (A^I, \overline A^{\overline J})$ always enters the scalar potential 
in the form \eqref{pot-intro}, 
no matter the supersymmetry breaking pattern or no matter if the theory is gauged or not, 
is a positive sign for the generality of our findings.

Finally, 
one of the most pressing questions concerns the string theory/brane origin of these new terms. 
The term in \cite{Cribiori:2017laj} is apparently related to the effective theory of the anti-D3 brane. 
The term we discuss here might have a similar origin, but we can only speculate on this at the moment.

\bigskip

\section*{Acknowledgments}

We thank Igor Bandos, Niccol{\`o} Cribiori, Ulf Lindstr\"om, Gabriele Tartaglino-Mazzucchelli and Antoine Van Proeyen for discussions. 
This work is supported from the KU Leuven C1 grant ZKD1118 C16/16/005.

\appendix

\end{document}